\documentstyle[prl,aps]{revtex}
\input{psfig}
\twocolumn
\begin{document}
\draft
\title{Solvable  glassy system: static versus dynamical transition}
\author{Th.~M.~Nieuwenhuizen}
\address{Van der Waals-Zeeman Laboratorium, Universiteit van Amsterdam
\\ Valckenierstraat 65, 1018 XE Amsterdam, The Netherlands}
\date{\today}
\maketitle
\begin{abstract}
A directed polymer is considered on a flat substrate  
with randomly located parallel ridges.
It prefers to lie inside wide regions between the ridges.
When the transversel width $W=\exp(\lambda L^{1/3})$ is exponential 
in the longitudinal length $L$, there can be a large number 
$\sim \exp L^{1/3}$ of available wide states. 
This ``complexity'' causes a phase transition from a
high temperature phase where the polymer lies in the widest lane, to
a glassy low temperature phase where it lies in one of many narrower lanes.

Starting from a uniform initial distribution of independent polymers,
equilibration up to some exponential time scale induces
a sharp dynamical transition.  
When the temperature is slowly increased with time, 
this occurs at a  tunable temperature. 
There is an asymmetry between cooling and heating.

The structure of phase space in the low temperature non-equilibrium glassy 
phase is of a one-level tree.
\end{abstract}
\pacs{0520,3620,6470P,7510N}
\narrowtext
The glass transition is caused by the appearance of a multitude
of long lived states, which prevents exploration of the whole
phase space. These effects are so strong that in practice one can
only observe precursor effects. 
Experimentally one observes a dynamical freezing around a tunable 
temperature.

The ergodic theorem says that time-averages may be replaced by
ensemble averages.
It is widely believed that the inherent dynamical nature of the 
glass transition implies that there is neither need nor chance 
for a thermodynamic explanation.  
However, since so many decades in time are involved,
 this is an unsatisfactory point of view.
J\"ackle~\cite{Jackle} and Palmer~\cite{Palmer} have therefore 
considered a glass as a system 
with many states (``components'') $a$,
each having its own component free energy $F_a(T)$. Within a state
thermodynamic equilibrium is established on moderate time scales,
while the system may escape to other states on much longer times.
It is often assumed that the states  occur
 with Gibbs weight $p_a= \exp(-\beta F_a(T))/Z$. From these assumptions
it was recently pointed out that the sharp dynamical glassy transition
of a typical mean field model, the $p$-spin interaction spherical 
spin glass, ~\cite{Ncomplexity}
can be explained thermodynamically: this transition is driven by the
complexity or configurational entropy 
(i.e. the logarithm of the number of relevant states). 

Here we wish to study this picture of the glassy transition in a simple, 
exactly solvable model without frustration,
both at the static and the dynamical level.

Consider  a directed polymer 
(or an interface without overhangs)
 $z(x)$ which lies in the region $1\le x \le L$ and
$1\le z \le W$ of the square lattice with unit lattice parameter.
In the Restricted Solid-on-Solid approximation  the interface
can locally be flat ($z(x+1)=z(x)$; no energy cost)
or make a single step ($z(x+1)-z(x)=\pm 1$; energy cost $J$).
The partition sum of this system, subject to periodic boundary conditions,
can be expressed in the $W \times W$ matrix ${\cal T}$ 
that transfers the system from $x$ to $x+1$
\begin{equation}
\label{Ztrace}
Z={\rm Tr}\, e^{-\beta{\cal H}}
={\rm Tr}\, {\cal T}^L=\sum_{w=1}^W \Lambda_w^L
\end{equation}
where $\Lambda_w$ is one of the eigenvalues of ${\cal T}$. 
For a pure system at temperature $T=1/\beta$
Fourier analysis gives the eigenvalues 
$\Lambda(k)= 1+2e^{-\beta J}\cos k$.
The largest eigenvalues occur at small momentum, 
$\Lambda(k)\approx \exp[-\beta f_B-\Gamma k^2/(2\pi^2)]$, 
which defines the bulk free energy density 
$f_B(T)=-T\ln(1+2e^{-\beta J})$  and the stiffness coefficient 
$\Gamma(T)=2\pi^2e^{-\beta J}/(1+2e^{-\beta J})$.

As depicted in Figure 1,
we shall consider the situation of randomly located potential barriers
parallel to the $x$-axis, so $V(x,z)=V(z)$.  
Hereto we assume binary disorder, $V(z)=0$ with probability
$p=\exp(-\mu)$ or $V(z)=V_1>0$ with probability $1-p$.
Eq. (\ref{Ztrace}) is dominated by the largest eigenvalues.
It is well known that they 
occur due to Lifshitz-Griffiths singularities.
~\cite{Lifshitz}~\cite{Griffiths}~\cite{NJPA}
They are lanes of width $\ell\gg 1$ in which all $V(z)=0$,
 bordered by regions with $V(z)\neq 0$. 
These dominant configurations are the 
``states'' or ``components'' of our system.
(In spin glass theory such states are  called ``TAP-states''.)
The eigenfunction centered around $z_a$
has  the approximate form $\cos[\pi (z-z_a)/\ell_a]$ inside the lane, 
while outside the lane it decays essentially  exponentially 
due to the disorder.
These states can thus be labeled by $a=(z_a,\ell_a)$. 
 Since $k\to \pi/\ell_a$ their free
  energy follows as
 $F_a=F_{\ell_a}$ where
\begin{equation}
\beta F_{\ell}\equiv -L\ln\Lambda(\frac{\pi}{\ell}) \approx \beta f_BL+
\frac{\Gamma L}{2\ell^2}
\end{equation}
The number of regions with $\ell$ successive sites with $V=0$ 
can be estimated by their ensemble average,
 ${\cal N}_\ell=W(1-p)^2p^\ell$. Since the relevant
$\ell$'s will turn out to be of order $L^{1/3}$, the interesting 
situation occurs when we choose $W=\exp(\lambda L^{1/3})$.
The states with width $\ell$ have a {\it configurational entropy}
or {\it  complexity}
${\cal I}_{\ell}\equiv \ln{\cal N}_\ell \approx \lambda L^{1/3}-\mu\ell$,
where $\mu=-\ln p$.
For large $L$ we may restrict the partition sum to these dominant
states. We thus evaluate, instead of eq. (\ref{Ztrace}), the ``TAP''
 partition sum
\begin{eqnarray}
Z= \sum_{\ell} {\cal N}_\ell e^{-\beta F_\ell} 
\end{eqnarray}
The total free energy 
\begin{equation} \label{bFtot}
\beta F=-\ln Z= L\beta f_B+\frac{\Gamma L}{2\ell^2}
-\lambda L^{1/3}+\mu\ell
\end{equation}
has to be optimized in $\ell$. 
The largest $\ell$ which occurs in the system can be estimated 
by setting ${\cal N}_\ell\approx 1 $,  yielding
\begin{equation} \ell_{max}=\frac{\lambda L^{1/3}}{\mu}
\end{equation}      
It is a geometrical length, independent of $T$.
Let us introduce $\gamma(T)=(\Gamma(T) \mu^2)^{1/3}/\lambda$.
The free energy of this widest state then reads
\begin{equation}
\label{Fliq}
\beta F=L \beta f_B+\frac{1}{2}\lambda L^{1/3}\gamma(T)^3
\end{equation}
It follows from (\ref{bFtot}) that at low enough $T$ 
the optimal width is smaller than $\ell_{max}$,
\begin{equation}
\ell^\ast=\left(\frac{\Gamma L}{\mu}\right)^{1/3}=\gamma(T) \ell_{max}
\end{equation}
The free energy of this phase is
\begin{equation} \label{bFopt}
\beta F =L\beta f_B+\frac{1}{2}\lambda L^{1/3}(3\gamma(T)-2)
\end{equation}
In the temperature interval where $\gamma(T)>1$ 
the interface lies in the non-degenerate widest lane.
For $\gamma(T)<1$ it lies in one of the ${\cal N}_{\ell^\ast}\gg 1$ 
optimal states,
each of which  has a higher free energy than the widest lane; this 
free energy loss is more than compensated by their complexity.
The system thus undergoes a static glassy transition at 
temperature $T_K=J/\ln(2\pi^2\mu^2\lambda^{-3}-2)$, where $\gamma=1$.
For any finite $L$ there is also a very low temperature regime 
$T<1/\log L$, where the interface is essentialy straight and can lie 
anywhere in the system, and a high temperature regime $T>\log L$,
where the potential barriers are ineffective, and the interface shape 
is truly random.

When considered as function of $\tilde T=1/T$ 
this situation is reminiscent of the $p$-spin interaction
spin-glass and to the Random Energy Model~\cite{Derrida}.
The very-high ${\tilde T}$ regime extends up to  ${\tilde T}\sim \ln L$,
 in which regime  a gradual freezing takes place in TAP
states of width $\ell^\ast$ much larger than unity.
This smeared transition is related to the sharp dynamical 
transition at some temperature $T_A>T_K$  of mean field spin-glass
models. In the regime ${\tilde T}_K<{\tilde T} \stackrel{<}{ \sim}\ln L$ 
 the system is frozen in TAP states of appropriate  degeneracy; 
a similar static phenomenon was found ~\cite{KTW}
for spin-glass models in the regime $T_K<T<T_A$.
Like in these models, 
below the ``Kauzmann'' temperature ${\tilde T}_K$ 
there is only an essentially non-degenerate state. This a manifestation of
the ``entropy crisis'' of glasses and glassy systems.

The internal energy of a state of width $\ell$ is
\begin{equation}
U_\ell=u_BL+\frac{L}{2\ell^2}\,\frac{\partial \Gamma}{\partial \beta}
=u_BL+\lambda L^{1/3}\frac{3}{2}
\frac{\partial \gamma}{\partial \beta}
\end{equation}
At the transition point $\gamma=1$ it is continuous, simply because
$\ell^\ast\to \ell_{max}$. 
The free energy (\ref{bFopt}) 
is a thermodynamic potential, and yields the same value for $U$.
There also holds the relation ${\rm d}U/{\rm d}T=T\,{\rm d}S/{\rm d}T$, 
where $S=S_{\ell^\ast(T)}+{\cal I}(T)=-{\rm d}F/{\rm d}T$ 
is the total entropy. 
On the side $\gamma<1$ the free energy (\ref{bFopt}) deviates quadratically 
from (\ref{Fliq}), leading to a higher specific heat. It reads
\begin{equation}
C=\frac{{\rm d}U}{{\rm d}T}=c_BL+\frac{L}{2\ell^2}
\partial_T\partial_\beta \Gamma+\frac{3}{\gamma}
\lambda L^{1/3} (T\partial_T \gamma)^2 
\end{equation}
It exceeds the component averaged specific heat of this phase,
${\overline C}=\sum_a p_a C_{\ell_a}
=Lc_B+(L/2\ell^2)\partial_T\partial_\beta \Gamma$.
The specific heat is larger in the glassy phase than in the 
paramagnet, because the free energy is lower. 
When considered as a function ${\tilde T}$, the specific heat makes a downward 
jump on cooling through ${\tilde T}_K$, as occurs in
realistic glasses.

When monitoring the internal energy of one polymer as a function of time,
as is easily done in a numerical experiment, one obtains essentially a noisy
telegraph signal. Each plateau describes trapping of  the polymer in one 
lane for some definite time. The variance of the noise in the internal energy
on this plateau is equal to $T^2\overline C$, implying 
fluctuations of order
$L^{1/2}$. From time to time the polymer moves to another lane, causing 
additional noise. The variance of the total noise equals $T^2C$, 
and it indeed exceeds $\overline C$ by an amount of order $L^{1/3}$.

We now consider dynamics.
On appropriate time scales,
our system can be viewed as a set of deep states (traps) $a$
located in lanes of width $\ell_a$ around centers $z_a$, 
with associated free energies $F_a\equiv F_{\ell_a}$ given in eq. (2). 
These minima are 
separated by very wide regions (separation $\sim \exp L^{1/3}$) with fully
random potential, that builds a barrier. Between traps $a$ and $a+1$ there 
is a free energy barrier determined by the intermediate state of highest
free energy. Let us call its  free energy $B_a$; it will typically lie 
at distance $L^{1/3}$ below the maximal free energy  
$Lf_{max}=-LT\ln\Lambda_{min}=-LT\ln(1-2e^{-\beta J})$~\cite{warning}.
The free energy barrier for the polymer to move from state
 $a$ to state ${a+1}$ is thus $B_a-F_{\ell_a}$, while the barrier for
moving  to the left is $B_{a-1}-F_{\ell_a}$.
For a statistical description of dynamics, we 
assume that many independent polymers are present, of which the units
 make random thermally activated moves.
On appropriate time scales one then gets a master equation 
for the probability $p_a(t)$ 
that the center of a polymer is inside the $a$'th state.
\begin{eqnarray}
t_0\frac {{\rm d} p_a(t)}{{\rm d} t}
&=&e^{\beta(F_{\ell_{a-1}}-B_{a-1})}p_{a-1}
+e^{\beta(F_{\ell_{a+1}}-B_{a})}p_{a+1}\nonumber\\
&-&e^{\beta(F_{\ell_a}-B_{a-1})}p_{a}
-e^{\beta(F_{\ell_a}-B_{a})}p_{a}
\end{eqnarray}
Here $t_0$ is the attempt time for a move of one  polymer segment.
This model is  a combination of the random jump-rate model and the random bond
models studied before.~\cite{HKL,NE}, see ~\cite{HK} for a review.
The stationary state is independent of the barriers
$B_a$:
\begin{equation}	
p_a^{eq}=\frac{e^{-\beta F_{\ell_a}}}{Z_{eq}}
\end{equation}
The denominator
$Z_{eq}$ is equal to the thermal TAP partition sum. So our master
equation exactly reproduces the Gibbs distribution discussed before.

Let us now consider the motion of polymers in an equilibrium ensemble,
after making some non-essential simplifications of the system.
 First we slightly modify the actual height of the barriers by setting
$B_a=Lf_{max}$,  thus neglecting their $L^{1/3}$ deviations.
Next we assume, for some fixed $\ell_0$ ($1\ll\ell_0\ll L^{1/3}$), 
that all relevant deep traps (and further a lot of shallow traps) 
are located at positions that are multiples of $W_{\ell_0}=\exp{\mu\ell_0}$, 
and we only consider those states.
Their number is ${\cal N}_{\ell_0}=\exp(\lambda L^{1/3}-\mu\ell_0)$.
Using a result of Haus et al, we know that  diffusive behavior is 
exact at all times,\cite{HKL}
\begin{equation} \label{zdiff}
\langle\delta z(t)^2\rangle\equiv \langle(z(t)-z(0))^2\rangle=D t
\end{equation}
involving the diffusion coefficient 
\begin{equation}  \label{D=}
D=\frac{2W_{\ell_0}^2 
{\cal N}_{\ell_0}}{t_0e^{\beta(Lf_{max}-F)}}
\end{equation}
with $F$ given in eq. (\ref{bFtot}). As the barriers have a  
height that deviates by order $L^{1/3}$ from a fixed value 
$L(f_{max}-f_{B})$,
we  introduce the logarithmic time variable $\tau$ 
~\cite{ell0scale} 
\begin{equation}\label{ttau}
t(\tau)=t_0
e^{\beta L(f_{max}-f_B)-\mu\ell_0}\times
e^{\lambda L^{1/3}\tau}
\end{equation} 
$\tau$ ranges from $\sim -L^{2/3}$, where $t\sim t_0$,
up to ${\cal O}(1)$, where $t\sim t_0\exp(L+L^{1/3})$
and the interesting physics occurs.

Let us now consider $T<T_K$.
 At given $\tau$ the polymers have a time $t(\tau)$ 
to make moves. The typical deviation follows from (\ref{zdiff}) 
and(\ref{D=}) as  $\log|\delta z|\sim \lambda L^{1/3}
(\tau+\ell/\ell_{max}+\gamma^3\ell_{max}^2/2\ell^2)/2$. 
This we can relate to the distance between optimal states 
$W_\ell=\exp\mu\ell$
and the number of them $\exp {\cal I}_{inter}(\tau)$, 
reached at time $t(\tau)$:
$|\delta z|\sim W_\ell \exp{{\cal I}_{inter}}$. 
It defines the {\it intercluster complexity},
i.e. the configurational entropy  of the states reached,
\begin{equation}\label{Is=}
{\cal I}_{inter}(\tau)=\frac{\lambda L^{1/3}}{2}
(\tau+\frac{\gamma^3\ell_{max}^2}{2\ell^2}-\frac{\ell}{\ell_{max}})
\end{equation}
The equilibrium relation $\ell=\gamma(T)\ell_{max}$ tells us that for 
$\tau=\gamma/2$ the next optimal state will typically have 
been reached,  so that the population of the states is in accord with the 
equilibrium prediction, and thermodynamic equilibrium is achieved.
We consider the bunch of $\exp{\cal I}_{inter}(\tau)$ states
 as one cluster.
There are $\exp{\cal I}_{intra}$ of these clusters, defining 
the {\it intra-cluster complexity}
\begin{equation}
{\cal I}_{intra}(\tau)={\cal I}-{\cal I}_{inter}(\tau)
=\lambda L^{1/3}(1-\frac{3}{4}\gamma-\frac{1}{2}\tau)
\end{equation}
As time progresses, more and more optimal states will be visited, 
expressed in the growth of ${\cal I}_{inter}$. This naturally 
occurs by a decrease 
of the number of independent clusters $\exp{\cal I}_{intra}$. 
At the ergodic timescale set by
\begin{equation}
\tau_{erg}(T)=\left\{{ 2-\frac{3}{2}\gamma(T) \qquad T<T_K \atop
                       1-\frac{1}{2}\gamma^3(T) \qquad T>T_K}
\right.
\end{equation}
ergodic behavior has been reached. Each polymer has been able to visit 
the whole system. The system is one big cluster
(${\cal I}_{intra}=0$) that contains all relevant states 
(${\cal I}_{inter}={\cal I}$).
Summarizing, so far we have found the regime B 
($0<T<T_K$; $\gamma(T)/2<\tau<\tau_{erg}(T)$),
where more and more optimal states are visited, and the regimes
C  ($T<T_K$; $\tau>\tau_{erg}(T))$ and D ( $T>T_K$; $\tau>\tau_{erg}(T))$
 of ergodic behavior. Thermodynamic equilibrium is
already reached in regime B, before the system is ergodic. 
Thermodynamic relations involving derivatives 
(e.g. ${\rm d}F/{\rm d}T$, ${\rm d}U/{\rm d}T$, ${\rm d}S/{\rm d}T$) 
are satisfied.

We now consider dynamics in the initial time regime A ($\tau<\gamma(T)/2$ 
for $T<T_K$ and $\tau<\tau_{erg}(T)$ for $T>T_K$).
Starting from a uniform initial distribution, where many narrow states are
populated, the polymers will move to wider and wider states.
At times $t(\tau)$ with $\tau<\gamma/2$
a polymer will visit a region of width $\delta z(\tau)$
where it picks up the broadest state available. 
The non-degeneracy of this state $({\cal I}_{inter}(\tau)=0)$ yields, 
now using eq. (\ref{Is=}) as an estimate, 
the best width reached up to then.
The complexity of these clusters is  ${\cal I}_{intra}
=\lambda L^{1/3}-\mu\ell(\tau)$.
This behavior may be expressed in terms of a dynamical partition sum
at time scale $t(\tau)$, where the system is split up in independent clusters 
$c=1,\cdots,{\cal N}_{\ell(\tau)}$ of width $W_{\ell(\tau)}$:
 $Z(\tau)=\sum_{c=1}^{{\cal N}_{\ell(\tau)}}Z_c(W_{\ell(\tau)})$
where each of the $Z_c(W)$'s is as in eq. (1).
This results in the  dynamical free energy 
\begin{eqnarray}
\beta F_{dyn}(\tau)= L\beta f_B+\lambda L^{1/3}
(\frac{\gamma^3\ell_{max}^2}{2\ell^2}
-1+\frac{\ell}{\ell_{max}})
\end{eqnarray}
Let us consider $T<T_K$. The dynamical free energy
has a minimum, which is approached in the limit $\tau\to\gamma(T)/2$.
The value of the minimum
coincides with the static value (\ref{bFopt}). At that timescale
a polymer will typically have found a 
state of optimal width $\ell(\tau)=\gamma\ell_{max}$. 
The behavior at larger times, $\tau>\gamma/2$, was already 
discussed above, and matches nicely with this short time dynamics.

In order to compare with cooling experiments in realistic glasses, 
we must consider a heating experiment (which is a cooling experiment
in the variable $\tilde T$). The temperature 
changes slowly with time, $T=T(\tau)$. It defines the inverse 
function $\tau(T)$, that characterizes the heating trajectory. 
 Due to the $L$-dependence in (\ref{ttau}),
$\tau$ will start at $\sim -L^{2/3}$ for small $t$, but when 
$\tau={\cal O}(1)$, it need not be a monotonically increasing function 
of $T$. Approaching $T_K$ from below under appropriate conditions, 
 a dynamical transition will occur at temperature $T_f<T_K$
(it is a freezing transition in terms of $\tilde T$). This temperature
is set by $\tau(T_f)=\gamma(T_f)/2$, where, starting from  small widths,  
the dynamically achieved width equals the optimal width, viz. 
$\ell(\tau(T_f))=\gamma(T_f)\ell_{max}$. At $T_f$ the internal energy is
continuous. For $T<T_f$ the specific heat reads 
\begin{equation} 
C=c_BL+\lambda L^{1/3}\left (
\frac{3\gamma^2}{2\ell^2}\partial_T\partial_\beta\gamma
+3\frac{\gamma^2|\partial_\beta\gamma|}{\ell^2}\partial_T
\ln\frac{\ell}{\gamma} \right) 
\end{equation}
For $T(t)>T_f$ the width $\ell$ will be equal to
the optimal value $\gamma(T(t))\ell_{max}$ 
and the last term vanishes. In the initial time regime,
where $\ell<\gamma$, this term  is positive. 
At $T_f$ the specific heat thus performs a jump, 
with the higher plateau value on the low temperature side, as occurs for
the static behavior at $T_K$. The height of this jump depends on the 
heating rate, and it vanishes for small enough rate. 
For large $L^{1/3}$ the jump is sharp in the logarithmic time 
variable $\tau$. 

In the dynamical regime A thermodynamic relations involving addition,
such as $F=U-TS$ remain valid, as they still arise from saddle point
analysis. But $F$ is not a thermodynamic potential.
Derivative relations, such as ${\rm d} \beta F/{\rm d} \beta =U$ 
and ${\rm d} S/{\rm d} T =C/T$ are violated. 

There is no symmetry between cooling and 
heating experiments. In order to have a decreasing $T(t)$ 
eq. (\ref{ttau}) tells us that $\tau$ must be of order $-L^{2/3}$,
whereas it is typically of order unity for heating.
Equating (\ref{Is=}) to  zero then leads to widths 
$\ell\sim {\cal O}(L^0)$, thus
completely reinitializing the relaxation. Such a phenomenon
was observed upon heating in spin glasses and explained
in terms of hierarchy of phase space.~\cite{Lefloch}

Our dynamical analysis puts forward the picture 
that the hierarchical structure of phase space,
here having the structure of a one level tree and
reminiscent of one step replica symmetry breaking, is a dynamical effect.
At given timescale only  nearby states can be reached, so there are
many clusters, each having one state (regime A) or many states (regime B). 
At larger times other states can be reached, thus leading to larger clusters. 
For times larger than the ergodic time of a large but finite system
(regimes C and D), all states are within reach, 
i.e. there is one cluster only, containing many states (regime C) or
one state (regime D).
In the thermodynamic limit
$L\to\infty$ before $t\to\infty$, 
phase space splits up in truly disjoint sets.

In contrast to most realistic glasses, our model has no crystal state.
Further study of the dynamics should show whether stretched exponential decay and
the Vogel-Fulcher law for the relaxation time occur.

\subsection*{ Acknowledgments}
The author thanks J.P. Bouchaud, J.M.F. Gunn, D. Lancaster, J.M. Luck,
 F. Ritort, G. Parisi, G.W. Wegdam, and Y.C. Zhang,
for stimulating discussion.

\begin{figure}[tbh]
\label{fig1}
\centering
\psfig{file=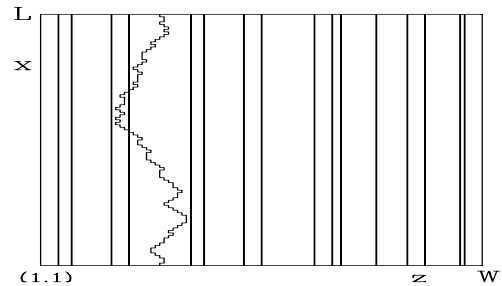,height=4cm,width=7cm}
\caption{A directed polymer can move on a substrate with
parallel potential barriers. It prefers to lie in
wide lanes between the barriers.}
\end{figure}

\begin{figure}[tbh]
\label{fig2}
\centering
\psfig{file=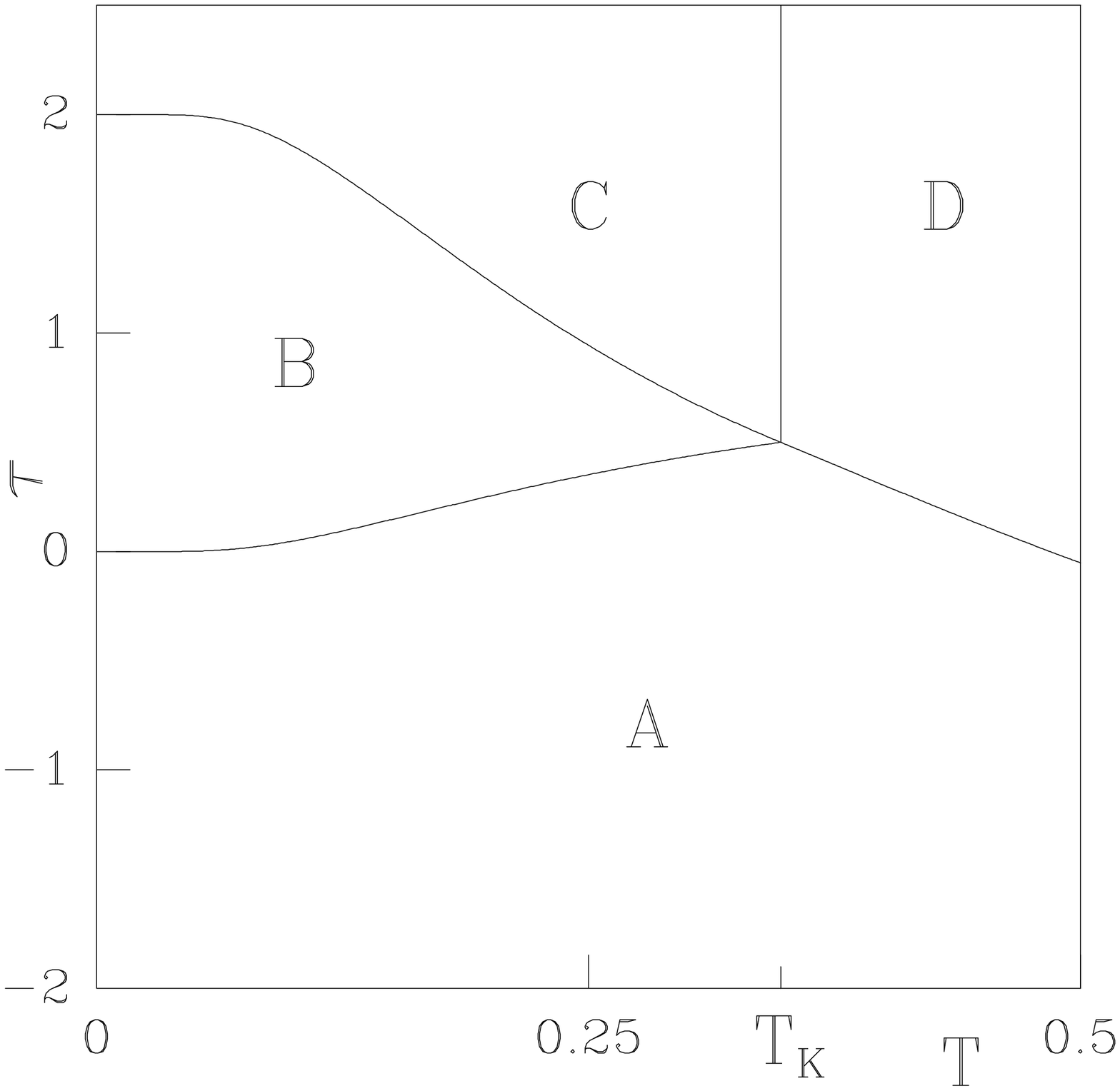,height=6cm,width=6cm}
\caption{
Dynamical phase diagram for $\lambda=\mu=J=1$. 
$\tau=c_1+c_2\ln t$ is logarithmic in time, $T$ is temperature and
$T_K$ the static Kauzmann transition temperature. 
A: initial regime; B: glassy regime with many states at 
thermodynamic equilibrium; 
C: ergodic glassy regime with many states; D: ergodic regime with one state.
When heating the system from short times (large negative $\tau$) 
and small $T$ towards larger $\tau$ and $T$, a dynamical (or ``experimental'') 
transition occurs at $T=T_g$ where the A-B or A-D boundary is crossed.
Crossing the
C-D boundary is only possible in the ergodic regime; it induces a 
static ``Kauzmann''-transition.}

\end{figure}

\end{document}